\documentclass[aps,pra,twocolumn,showpacs]{revtex4}
\usepackage[dvips]{graphicx}
\usepackage{amsmath}
\usepackage[latin1]{inputenc}
\usepackage[frenchb]{babel}

\begin{document}

\title{Diffraction limited optics for single atom manipulation}

\author{Y.R.P.~Sortais\footnote{Electronic address : yvan.sortais@institutoptique.fr}, H.~Marion, C.~Tuchendler, A.M.~Lance,
M.~Lamare, P.~Fournet, C.~Armellin, R.~Mercier, G.~Messin, A.~Browaeys and P.~Grangier}
\affiliation{Laboratoire Charles Fabry de l'Institut d'Optique, Campus Polytechnique, RD 128, 91127
Palaiseau Cedex, France} \altaffiliation{Laboratoire Charles Fabry is an ``Unit\'e Mixte de
Recherche" of Institut d'Optique Graduate School, Centre National de la Recherche Scientifique, and
Universit\'e Paris-Sud.}

\date{\today}

\begin{abstract}
We present an optical system designed to capture and observe a single neutral atom in an optical dipole trap, created by focussing a laser beam using
a large numerical aperture ($\rm{N.A.}=0.5$) aspheric lens. We experimentally evaluate the performance of the optical system and show that it is
diffraction limited over a broad spectral range ($\sim200$~nm) with a large transverse field ($\pm25~\mu$m). The optical tweezer created at the focal
point of the lens is able to trap single atoms of $^{87}$Rb and to detect them individually with a large collection efficiency. We measure the
oscillation frequency of the atom in the dipole trap, and use this value as an independent determination of the waist of the optical tweezer.
Finally, we produce with the same lens two dipole traps separated by $2.2~\mu$m and show that the imaging system can resolve the two atoms.
\end{abstract}


\maketitle
\section{Introduction}
The observation and manipulation of a few individual particles are at the core of many present experiments in atomic and molecular physics, quantum
optics, quantum information, as well as in biology and chemistry. Quite often, these experiments rely on high numerical aperture optics which collect
a very weak fluorescence signal emitted by the particles. These optics generally operate at the diffraction limit,  in order to image small objects
with a high spatial resolution. As a few examples, the objects may be single ions in a Paul trap~\cite{Dehmelt1980}, single neutral atoms in
microscopic optical dipole traps~\cite{Schlosser2001}, Bose-Einstein condensates in a double well potential~\cite{Albiez2005}, or single fluorophores
in a biological membrane~\cite{Wenger2006}. The high resolution of the imaging optics allows the observation of periodic chains of
ions~\cite{Roos2004,Wineland} or arrays~\cite{Bergamini2004} of microscopic objects, with distances between them as small as a few microns.

Many of these experiments rely on the ability not only to observe,  but also to trap and manipulate the  particles. Large numerical aperture optics,
when diffraction limited, can focus laser beams down to sub-micron spots that can be used as sharp optical tweezers~\cite{Grier2003}. In atomic
physics, this strong confinement can be used to trap exactly one atom in the tweezer~\cite{Weber2006,Schlosser2001}, with high oscillation
frequencies due to the sharp focussing. Using independently controlled optical tweezers simultaneously, one can control the collision between two (or
more) individual particles~\cite{Calarco2000}. This approach has already been implemented to investigate interactions between moving biological
objects~\cite{Mammen1996}, and offers an interesting perspective to realize quantum logic operations between few cold neutral atoms or
ions~\cite{Dorner2005,Roos2004}.

In the case of atoms or ions, the design of large numerical aperture optics requires to take into account the ultra-high vacuum environment that is
necessary to produce and manipulate them. The optics may be either inside the vacuum chamber - and then must be bakable and vacuum compatible - or
outside - but then the focussed beam must go through a vacuum window, which generally creates significant aberrations. Another constraint arises from
the arrangement of the trapping system surrounding the particles (e.g. electrodes for an ion trap or laser cooling beams), which usually requires a
long enough working distance of at least a few millimeters. These constraints add up to the requirements of large numerical aperture and diffraction
limited performance, 
and often lead to a rather complicated design and manufacturing of vacuum compatible custom objectives.

In this article, we describe and characterize a simple optical system, based on the combination of a large numerical aperture aspheric lens placed
inside the vacuum chamber, and a few standard lenses placed outside. This system combines the powerful techniques of optical tweezers and confocal
microscopy to trap, manipulate and observe single ultra-cold $^{87}$Rb atoms. The simplicity and low cost of the design compare favorably with custom
objectives based on spherical lenses~\cite{Vigneron1998,Alt2002} and suggest broad applicability to other fields of research where excellent spatial
resolution is critical. The numerical aperture is $\textrm{N.A.} = 0.5$ and, for a fixed working distance on the order of a centimeter, it performs
at diffraction limit over a large spectral range, from 700~nm up to 880~nm~\cite{numerical_aperture}. We take advantage of this property to both
focus the tweezer beam at a wavelength of 850 nm,
and collect  fluorescence light at 780~nm through the same lens.

The paper is organized as follows. Section~\ref{section:requirements} details  the requirements of the system and describes the optical setup. In
Section~\ref{section:optical_characterization} the performance of the tweezer is characterized using optical techniques.
Section~\ref{section:single_atom_detection} explains how to trap and detect a single atom at the focal point of the tweezer, and
Section~\ref{section:oscillation_frequency} how to  measure the oscillation frequency of one atom in the trap. This method provides a way to probe
the light field locally, and to perform an independent measurement of the laser beam waist at the focal spot.

\section{Requirements and design of the optical system}\label{section:requirements}
Trapping a single atom in an optical tweezer leads to the following requirements for our objective lens. First, controlling the evolution of the
atom, or its interactions with another atom, requires that the trap size be smaller than a few microns, i.e. that the objective be diffraction
limited while having a numerical aperture as large as possible. In particular, our method for single atom trapping relies on a ``collisional blockade
mechanism"~\cite{Schlosser2002}: in a very small trap, the two-body loss rate is so high that if an atom is already trapped, a second atom entering
the trap leads to a fast inelastic collision and, eventually, to the loss of the two atoms. Second, loading of the tweezer is performed by focussing
the laser beam into a reservoir of laser cooled atoms, an optical molasses in our case. In practice, this implies that the focussing lens has a large
enough working distance to allow for optical access of the cooling beams, typically at least 5 to 10~mm is desirable. Third, the laser used to
produce the optical tweezer is far off-resonance at 850~nm, in order to avoid heating and decoherence of the trapped atom due to spontaneous
emission. On the other hand, the atom is probed by exciting the D${_2}$ transition and collecting the fluorescence at 780~nm, with a dichroic plate
separating the two radiations. Since it is convenient to have both beams going through the same aspheric lens, a third requirement is that the system
be diffraction limited over a broad spectral range while keeping the working distance constant \cite{achro}. Fourth, in view of future experiments
using several single atoms trapped in adjacent tweezers, it is also desirable that the objective remains diffraction limited off axis, which requires
a large enough transverse field. For instance, a field of view of  $\pm25~\mu$m and a resolution of  2~$\mu$m (see below) allows one to easily
address several hundreds of traps. Designing arbitrary arrays could be achieved for example by combining our large N.A. lens and a spatial phase
modulator placed on the incoming laser beam~\cite{Bergamini2004}, or by using appropriate optical lattices~\cite{Peil2003}.

The requirements enumerated above can be fulfilled by combining a sufficient number of spherical lenses. This is the case for the microscope
objective that we designed for our first generation experiment dedicated to single atom manipulation~\cite{Vigneron1998,Darquie2005}. This objective
consists of nine spherical lenses and has a numerical aperture $\rm{N.A.}=0.7$. Tight mechanical constraints and careful alignment of the lenses
resulted in a diffraction limited performance with spatial resolution of $0.7~\mu$m. The transverse field is $\pm10$~$\mu$m. The working distance is
10~mm and the objective operates under ultra-high vacuum.

For simplification and scalability purposes, we have built a second generation apparatus that uses a single commercial molded aspheric
lens~\cite{LightPath_reference} with $\rm{N.A.}=0.5$, a working distance of $\sim 5.7$~mm and a focal length of 8~mm. This lens is manufactured by
LightPath Technologies,~Inc.~\cite{IOTA_non_endorsment} and is diffraction limited at 780~nm for an object plane at infinite distance, with a
$0.25$~mm thick glass plate between the lens and the image focal plane. The absence of this glass plate results in an enhanced spherical aberration
when operating the lens at an infinite conjugate ratio. However, we noticed that aberration-free operation can be restored by slightly defocussing the
aspheric lens, and operating with weakly non-collimated beams. Low numerical aperture lenses can provide such beams without introducing aberrations.
This beam shaping, which is crucial for our system, was optimized by using an optical design software~\cite{Code_V}.
\begin{figure}
\includegraphics[width=8cm]{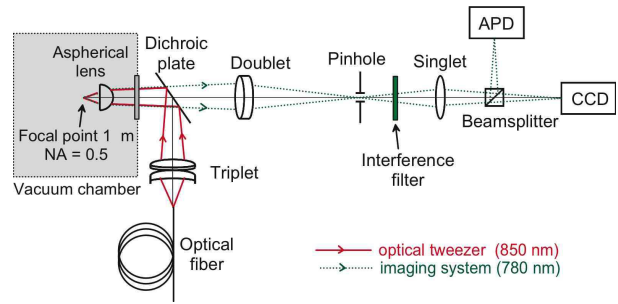}
\caption{(Color Online) Optical set-up of our trapping (solid line) and imaging (dotted  lines) systems. For clarity, schematic is not to scale.}
\label{Schema_optique_aspherix}
\end{figure}

The optical setup is shown in Fig.~\ref{Schema_optique_aspherix}. The source used for the tweezer is a single mode laser diode operating at 850~nm.
We focus this light into a single mode polarization maintaining fiber with an output $\textrm{N.A.}=0.11$. The beam at the output of the fiber is
shaped by using a triplet~\cite{Melles_Griot_reference1} outside the vacuum chamber that provides a slightly converging beam. Then it is injected
into the vacuum chamber and onto the lens mounted under ultra-high vacuum~\cite{footnote_lens_baking}. The same aspheric lens is then used as an
imaging system, to collect the fluorescence emitted by the atoms trapped in the optical tweezer.
The collection solid angle is $\Omega/4 \pi = 0.076$. A polarization independent dichroic plate separates fluorescence light at 780~nm from the tweezer
light at 850~nm. Imaging of the atomic samples is performed by using a doublet and a singlet~\cite{Melles_Griot_reference2} outside the vacuum
chamber. The overall transverse magnification of the imaging system is $\sim 25$, which allows imaging of our 1~$\mu$m diameter dipole trap onto a
2~pixels$\times2$~pixels area of a CCD camera (pixel size is 13~$\mu$m$\times$13~$\mu$m). We also use a polarization beam-splitter to collect part of
the fluorescence onto an avalanche photodiode (APD).

Once the working distance between the aspheric lens and the focal point has been fixed, the distances between the lenses shown in
Fig.~\ref{Schema_optique_aspherix} can be varied along the optical axis by a few centimeters, whilst keeping the system diffraction limited.
Likewise, the transverse alignment of the lenses and the fiber source is tolerant within $\sim1$~mm.

\section{Optical characterization of the tweezer}\label{section:optical_characterization}
In practice, aberrations of the optical system deform the wave front and result in energy being spread away from the central lobe of the diffraction
image, thus leading to a peak intensity attenuation and  to a shallower dipole trap. We quantify the amount of aberrations of the
optical system by considering the ratio of the peak intensity $I_{\rm{aberr}}$ in presence of aberrations to the intensity $I_{\rm{stig}}$ which
would be obtained if the optical system were perfectly stigmatic (i.e. free of aberrations). This ratio is called the Strehl ratio~\cite{Born_Wolf}
and is related to the deformation of the wave front as follows:
\begin{eqnarray*}\label{eq:Strehl_ratio}
S=\frac{I_{\rm{aberr}}}{I_{\rm{stig}}}\simeq 1-4\pi^2\frac{\Delta^2}{\lambda^2}
\end{eqnarray*}
Here, $\Delta$ denotes the root-mean-square departure of the actual wave front with respect to the ideal one, and $\lambda$ is the wavelength of the
radiation propagating through the system. A practical criteria is that the peak intensity attenuation $S$ is larger than the arbitrary value $0.8$
for the amount of aberrations to be acceptable ($S\geq0.8$). This sets an upper limit to the amount of aberrations that
we tolerate in our system ($\Delta\leq\lambda/14$). For a system free of aberrations, $\Delta=0$ and $S=1$.
Experimentally, the optical tweezer was characterized in three steps.

First, we tested the performance of the triplet and the aspheric lens
separately by using a wavefront analyzer (Fizeau interferometer~\cite{Born_Wolf}). We find that both exhibit diffraction limited performance:
$\Delta\leq\lambda/40$ for the triplet ($\rm{N.A.}=0.12$) and $\Delta\leq\lambda/30$ for the aspheric lens ($\rm{N.A.}=0.5$), the latter being
remarkable for such a large numerical aperture. The residual deformation of the wave front is due mainly to a small spherical aberration.

\begin{figure}
\includegraphics[width=8cm]{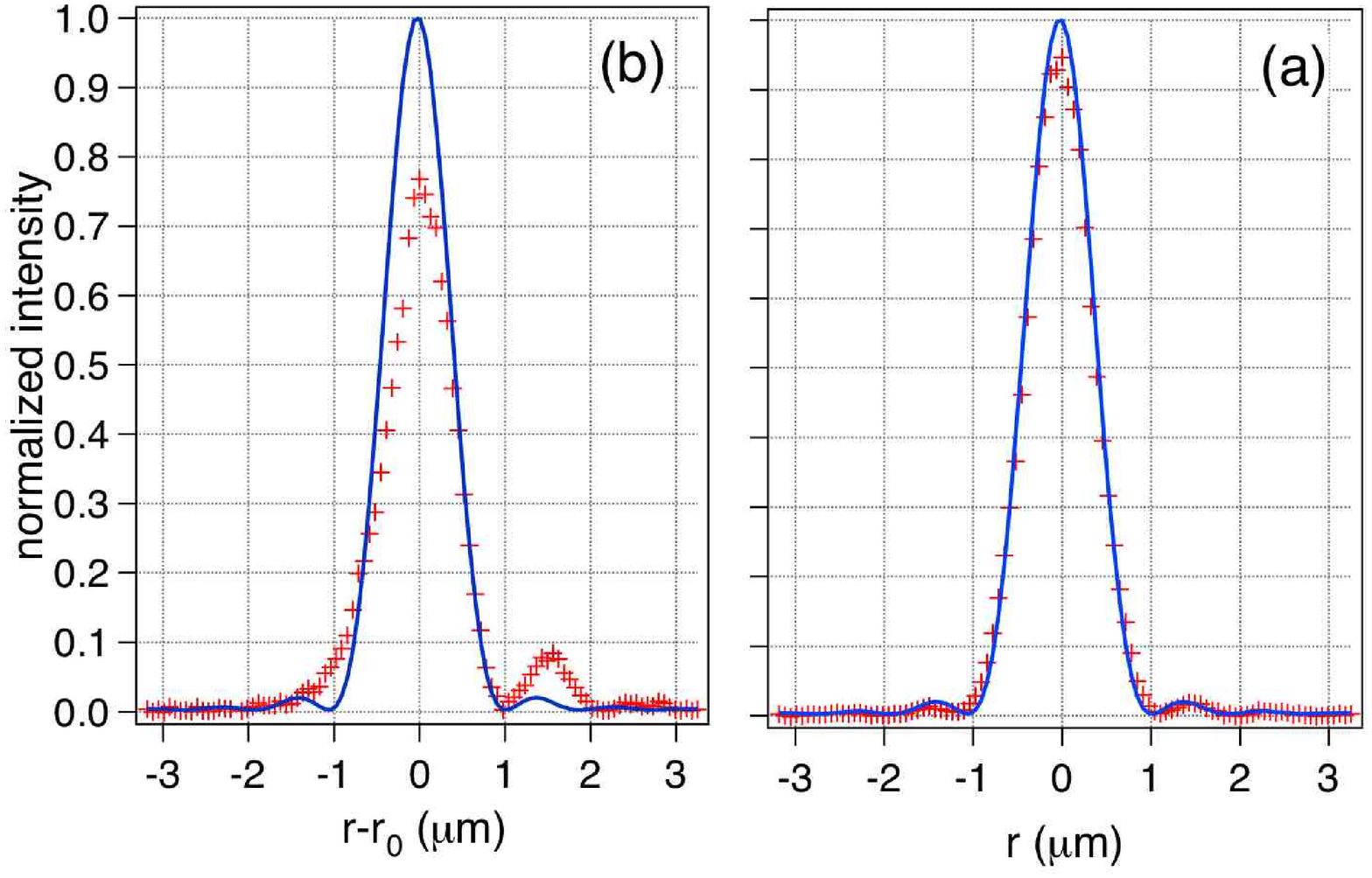}
\caption{(Color Online) (a) Cross-section of the point spread function of the tweezer system, measured at 850~nm on axis (crosses). We show the Airy
function (solid line) for comparison. The measured peak intensity is $S=0.93$ on axis. When the point source is moved off-axis by $r_0=30~\mu$m, the
point spread function gets distorted and the peak intensity is attenuated ($S=0.77$). (b) Cross-section along the direction of transverse
displacement of the point source (crosses) and the Airy function (solid line). Total flux is kept constant in both cases.} \label{PSFcut}
\includegraphics[width=8cm]{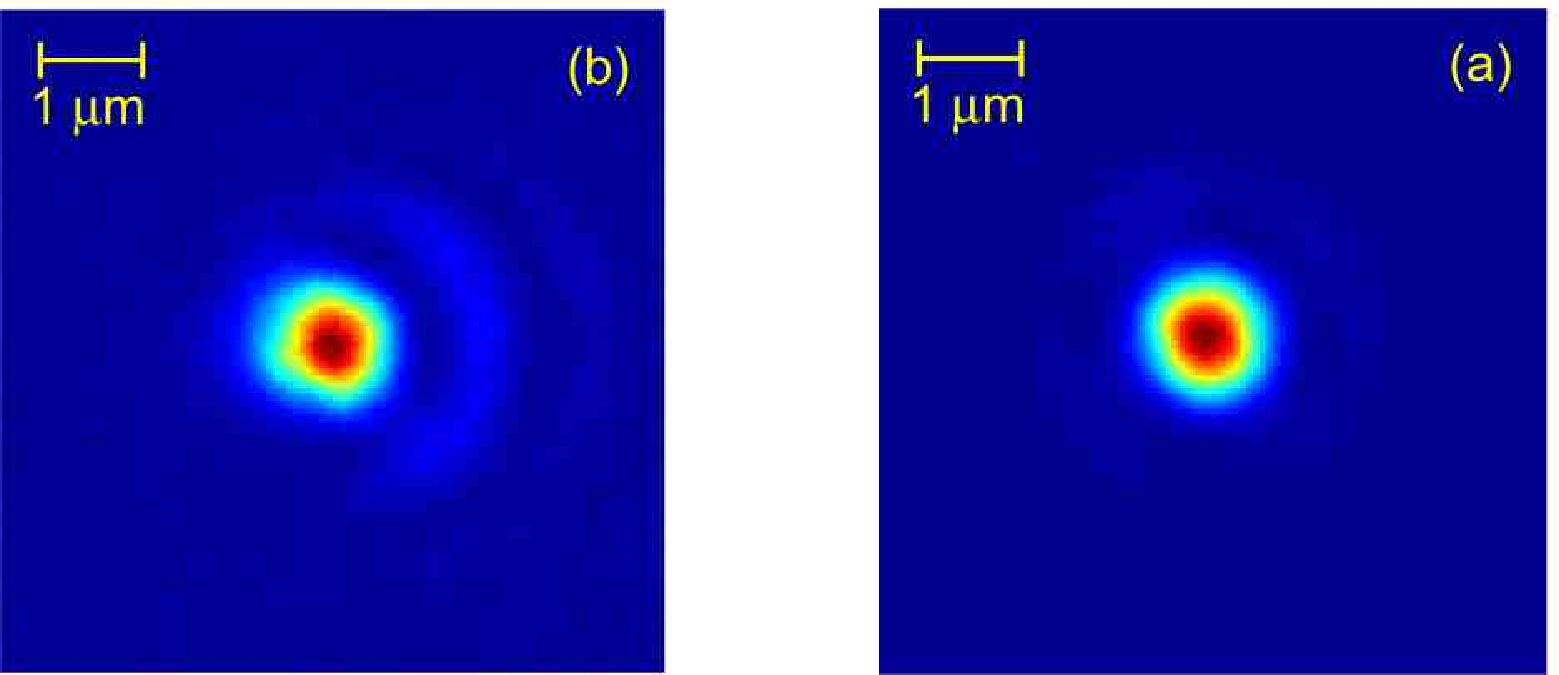}
\caption{(Color Online) Images of the point spread function of our tweezer system, taken with a linear CCD camera at 850~nm, for a sub-micron pinhole
source on axis (a) and off-axis by 30~$\mu m$ (b). Cross-sections shown on Fig.~\ref{PSFcut} were taken along the horizontal axis of these images,
across the center, where intensity is maximum.} \label{PSFimages}
\end{figure}

\begin{figure}
\includegraphics[width=8cm]{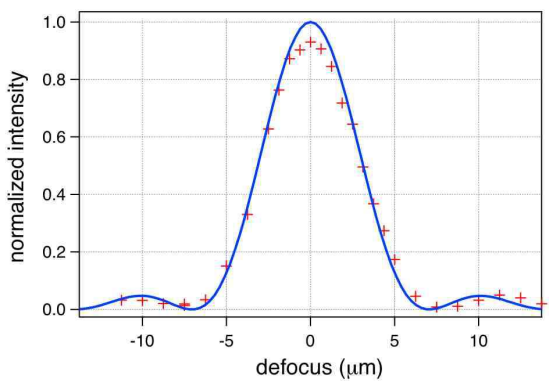}
\caption{(Color Online) On-axis intensity variation close to the focal point, as a function of the position along the optical axis. The prediction
from diffraction theory is shown for comparison, with the intensity normalized as in Fig.~\ref{PSFcut}. } \label{onaxis}
\end{figure}

For the second step in characterizing the optical tweezer, we aligned the triplet and the aspheric lens and evaluated the Strehl ratio $S$ of the
ensemble by analyzing point spread functions of the system and comparing them to the ideal case, i.e. with no aberrations present. In order to record
the point spread function, we used a back-illuminated sub-micron pinhole in the focal plane of the aspheric lens, in place of the cold atoms. The
pinhole being much smaller than $1.22 \; \lambda/\rm{N.A.}=2$~$\mu$m, its intensity diagram is flat over the full aperture of the aspheric lens. The
optical response of the system to this point source illumination was observed on a linear 16~bits CCD camera after magnification by a factor $50$ by
an aberration-free objective. Figure~\ref{PSFcut}(a) shows a cross-section of the point spread function measured on axis in the best focus plane,
corrected for the $(\times 50)$ magnification of the observation objective and for the $(\times 4)$ magnification of the $(\textrm{triplet, aspheric
lens})$ system. The corresponding CCD image is shown on Fig.~\ref{PSFimages}(a). Reversal of light propagation ensures that this actually is the
point spread function of the system, had it been illuminated by an isotropic point source in the focal plane of the triplet
(see~Fig.~\ref{Schema_optique_aspherix}). It should be compared to the $(2J_1(\zeta)/\zeta)^2$ Airy function predicted by diffraction theory. Here
$\zeta=2\pi r\times \rm{N.A.}/\lambda$, where $r$ is the radial coordinate, $\rm{N.A.}=0.5$, and $\lambda=850$~nm. The full width at half maximum of
the measured point spread function is $\textrm{FWHM}_\perp=0.9~\mu$m and the radius of the first dark ring is 1.06~$\mu$m, which is in agreement with
diffraction theory within a few percent. Normalizing the peak intensity of the Airy function to 1, we measure a peak intensity of $0.93$ for our
system (the total flux is kept constant in both cases), which is a measure of the Strehl ratio $S$ on axis. For a full characterization of the light
potential that the atomic samples will explore, we also scanned the magnifying objective along the axis around the best focus position.
Figure~\ref{onaxis} shows the measured on-axis intensity variation, corrected for the longitudinal magnification of the objective and
$(\textrm{triplet, aspheric lens})$ system. The on-axis intensity vanishes symmetrically $7~\mu$m away from the best focus position, and bright rings
appear off-axis, in place of the previous dark rings (contrast inversion). The full width at half maximum of the longitudinal intensity distribution
is $\textrm{FWHM}_\parallel=6.3~\mu$m, which again is in good agreement with diffraction theory.

\begin{figure}[b]
\includegraphics[width=8cm]{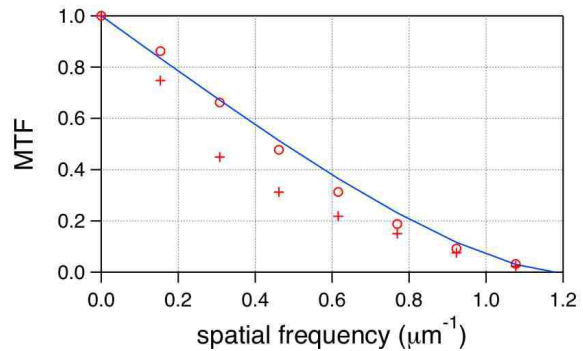}
\caption{(Color Online) Modulation transfer function (MTF) of the tweezer system. MTF measured on-axis (circles) and off-axis by 30~$\mu$m (crosses)
are compared to the theoretical frequency response of an aberration-free system (solid line). Strehl ratio is $S=0.93$ on-axis, and $S=0.77$
off-axis.} \label{FTM_vs_field}
\end{figure}

Our final step in characterizing our optical tweezer was investigating the off-axis performance of our system by moving the pinhole perpendicular to
the optical axis. Figure~\ref{PSFimages} shows the point spread functions measured for a pinhole respectively on-axis and off-axis by $30~\mu$m. The
latter displays the characteristic V-shaped flare of a comatic aberration. Since our system is nearly free of aberrations on axis, we determine its
Strehl ratio by directly comparing the peak intensities of images taken with an off-axis and on-axis pinhole, provided the total flux is kept
constant [see Fig.~\ref{PSFcut}(b)]. The performance of the optical tweezer can be equivalently evaluated by plotting its optical modulation transfer
function (MTF) that characterizes the attenuation of the various spatial frequencies present in a test object, due to the optics~\cite{MTF}.
Figure~\ref{FTM_vs_field} shows cross-sections of the 2D-Fourier transform of the images shown in Fig.~\ref{PSFimages}, and compares them to the
frequency response of an aberration-free system with similar numerical aperture: the 2D-autocorrelation function of a circular aperture. The
measurements show that our system displays a field of $\pm25~\mu$m over which $S\geq0.8$.

It should be noted here that the results described above were obtained by illuminating the system with spherical wavefronts, issued from a sub-micron
pinhole. The situation is slightly different for our single atom tweezer, because the system is illuminated with a gaussian beam provided by an
optical fiber, with a beam waist equal to the aspheric lens radius. This has two consequences~: first, the diffraction rings structure shown above is
significantly damped, due to the gaussian apodization effect. Second, the  point spread function becomes slightly broader, with a $\rm{FWHM}_\perp$
increased by $9\%$, and is neither an Airy function nor purely gaussian.

\section{Detection of a single atom trapped in an optical tweezer}\label{section:single_atom_detection}
The optical tweezer described above is a powerful tool for manipulating single neutral atoms. The alignment of the optical tweezer onto the center
region of an optical molasses of cold $^{87}$Rb atoms indeed results in the trapping of a single atom in the microscopic dipole trap as shown below.
The optical molasses~\cite{Lett1989,Dalibard1989} is produced by six counter-propagating cooling beams at $\sim780$~nm and a Zeeman slowed beam of Rb
atoms~\cite{Phillips1982}, providing a large reservoir of cold atoms (with typical size $\sim 1$~mm) surrounding the microscopic optical tweezer.
\begin{figure}[b]
\includegraphics[width=8cm]{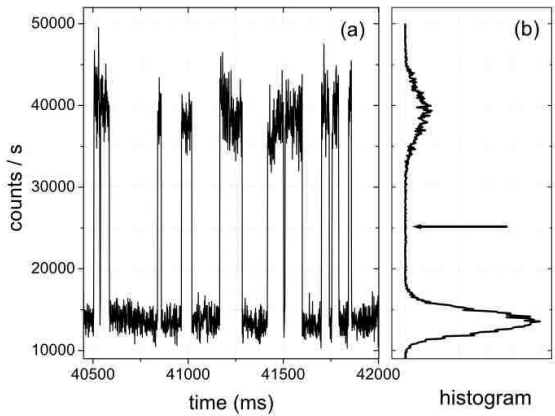}
\caption{(a) Fluorescence of a single atom measured by the APD. Each point corresponds to a 10~ms time bin. (b) Histogram of the measured
fluorescence recorded over 100~s. The two Poisson distributions correspond to the presence and the absence of a single atom in the dipole trap. The
arrow indicates the threshold that we use to discriminate between the two cases.} \label{fluorescence_atome_unique}
\end{figure}

We observe the fluorescence of the trapped atoms with an avalanche photodiode (APD) used in single photon counting mode, as well as a CCD camera with
low read-out noise (see Fig.~\ref{Schema_optique_aspherix}). We use an interference filter (centered at 780~nm, with 10~nm bandwidth) and a
400~$\mu$m diameter pinhole in confocal configuration to reduce the background signal. The residual scattering  of the 780~nm cooling beams in the
vacuum chamber and on the optical table contributes for $81\%$ of this background signal, the remaining $19\%$ coming from the background
fluorescence at 780~nm by the molasses itself. We estimate the overall collection efficiency of our system to be $\sim1\%$, taking into account the
angular collection efficiency of the lens ($7.6\%$), the transmission of the optics and the quantum efficiency of the APD.

Figure~\ref{fluorescence_atome_unique}(a) shows a time sequence of the fluorescence signal collected on the APD observed in $10$~ms time bins.
For this figure, the molasses beams and the dipole trap are continuously on. The sudden jumps in the fluorescence signal correspond to a
single atom entering the optical tweezer, while the sudden drops of the fluorescence signal correspond to the atom leaving the trap due to an
inelastic collision with another entering atom. The collision process occurs because it is light-assisted by the molasses beams at 780~nm
(collisional blockade mechanism) and results in the two atoms being ejected from the trap. The absence of two-atom steps is by itself an indication
that the waist of the trapping beam is no more than a few microns, since this condition is required for the collisional blockade
mechanism~\cite{Schlosser2002}. In that case, the loss rate due to two-body collisions becomes huge when the trapping volume is very small, and pairs
of atoms are never observed.

For the data of Fig.~\ref{fluorescence_atome_unique}, we use 5.6~mW of trapping light at 850~nm. Assuming a waist of $1~\mu$m (see
Sec.~\ref{section:oscillation_frequency} below), we calculate a trap depth of $1.5$~mK, or equivalently a lightshift of $32$~MHz. The cooling beams
that induce the fluorescence of the atom are red-detuned with respect to the atomic transition by $5\Gamma$, where
$\Gamma=2\pi\times6\times10^6$~rad.s$^{-1}$ is the linewidth of the D$_2$ transition. The fluorescence signal exhibits steps of
$2.7\times10^4$~photons.s$^{-1}$ on top of a background signal of $1.3\times10^4$~photons.s$^{-1}$. The analysis of the corresponding histograms
shown in Fig.~\ref{fluorescence_atome_unique}(b) indicates that both background and step signals are shot-noise limited. This, together with the step
height of $2.7\times10^4$~photons.s$^{-1}$, allows us to discriminate between the presence and the absence of an atom within $10$~ms with a
confidence better than $99\%$. In the presence of the cooling laser beams, the storage time of the atom is limited by the light-assisted collision
with a second atom. It can be varied from 100~ms up to about 10~s, depending on the density of the molasses cloud.

We have also measured the lifetime of the atom in the absence of the cooling light and found a $1/e$ decay time of $\sim10$~s. This was measured by
varying the duration $T_{\rm{off}}$ during which the cooling beams are switched off, and measuring the probability to detect the atom fluorescence
again right after $T_{\rm{off}}$. This lifetime may be limited by collisions with the background gas (pressure in the $10^{-9}$~Pa range) as well as
heating mechanisms.
\begin{figure}
\includegraphics[width=9cm]{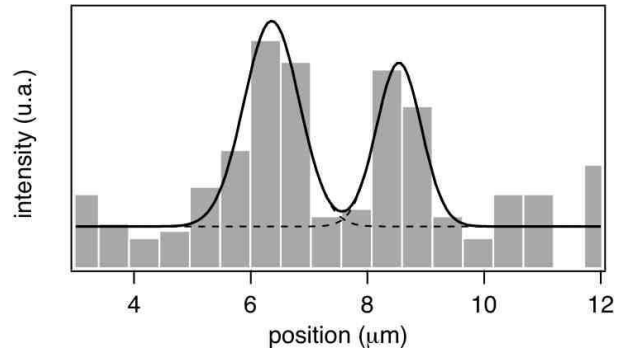}
\caption{Cross-section of a CCD image showing two single atoms trapped in two adjacent optical tweezers, corrected for the magnification ($\times25)$
of the imaging system. The distance between the two optical tweezers is $2.2\pm0.1~\mu$m. Each peak is fitted by a gaussian model (dashed lines) and
exhibits a waist $w=0.9\pm0.2~\mu$m. The solid line represents the sum of the fits of the two fluorescence signals emitted by each single atom.
Vertical bars represent the intensity measured by each pixel of the CCD camera during a time window of $100$~ms.} \label{doubleatomeunique}
\end{figure}

Finally, we briefly address the issue of addressability and resolution of our imaging system. For this purpose, we produce two tweezers by sending a
second trapping beam at 850~nm at a small angle in the aspheric lens. An angle of 0.25~mrad, together with the focal length of 8~mm of the aspheric
lens, results into a 2~$\mu$m distance between the two traps. The loading of the two traps is not deterministic but we can easily  find periods of
time during which two atoms are present at the same time in the two optical tweezers. Figure~\ref{doubleatomeunique} shows a cross-section of a CCD
image taken when such an event occurs. A gaussian fit to the peak signal produced by each single atom indicates a waist of $0.9\pm0.2~\mu$m, which
validates the performance of the tweezer system and the performance of the imaging system altogether. We emphasize that the two atoms are always
present during the 100~ms integration time. Therefore this data corresponds to realistic conditions for ``reading out" a quantum
register~\cite{Dotsenko2005} with two qubits separated by 2.2~$\mu$m. This indicates that our imaging system may resolve two atoms with separation as
small as $2~\mu$m, within a time as short as 10~ms.

\section{Measurement of the transverse oscillation frequency}\label{section:oscillation_frequency}
Once the atom is trapped, we switch off the cooling beams, whilst the optical tweezer is kept on. The atom oscillates at the bottom of the dipole
trap, and measurement of its oscillation frequency provides an \textit{in-situ} measurement of the trap dimension. Knowing the power
$P_{\textrm{trap}}$ of the trapping laser, and assuming that the trapping beam is gaussian, we can calculate from the oscillation frequency the waist
$w_0$ of the optical tweezer:
\begin{eqnarray*}
w_0=\left[\frac{\hbar \Gamma}{m \omega_r^2}\frac{P_{\textrm{trap}}}{\pi
I_{\textrm{sat}}}\left(\frac{\Gamma}{3\delta_1}+\frac{2\Gamma}{3\delta_2}\right)\right]^{1/4}
\end{eqnarray*}
where $I_{\textrm{sat}}\simeq1.67$~mW/cm$^{2}$ is the saturation intensity of the D$_2$ transition, $\omega_r$ is the oscillation
frequency of the atom measured in the radial plane (i.e. perpendicular to the tweezer optical axis), and
$\delta_1\simeq2\pi\times2.4\times10^{13}$~rad.s$^{-1}$ (resp. $\delta_2\simeq2\pi\times3.2\times10^{13}$~rad.s$^{-1}$) is the
frequency detuning of the tweezer beam relative to the D$_1$ (resp. D$_2$) transition~\cite{Grimm2000}.

\begin{figure}
\includegraphics[width=8cm]{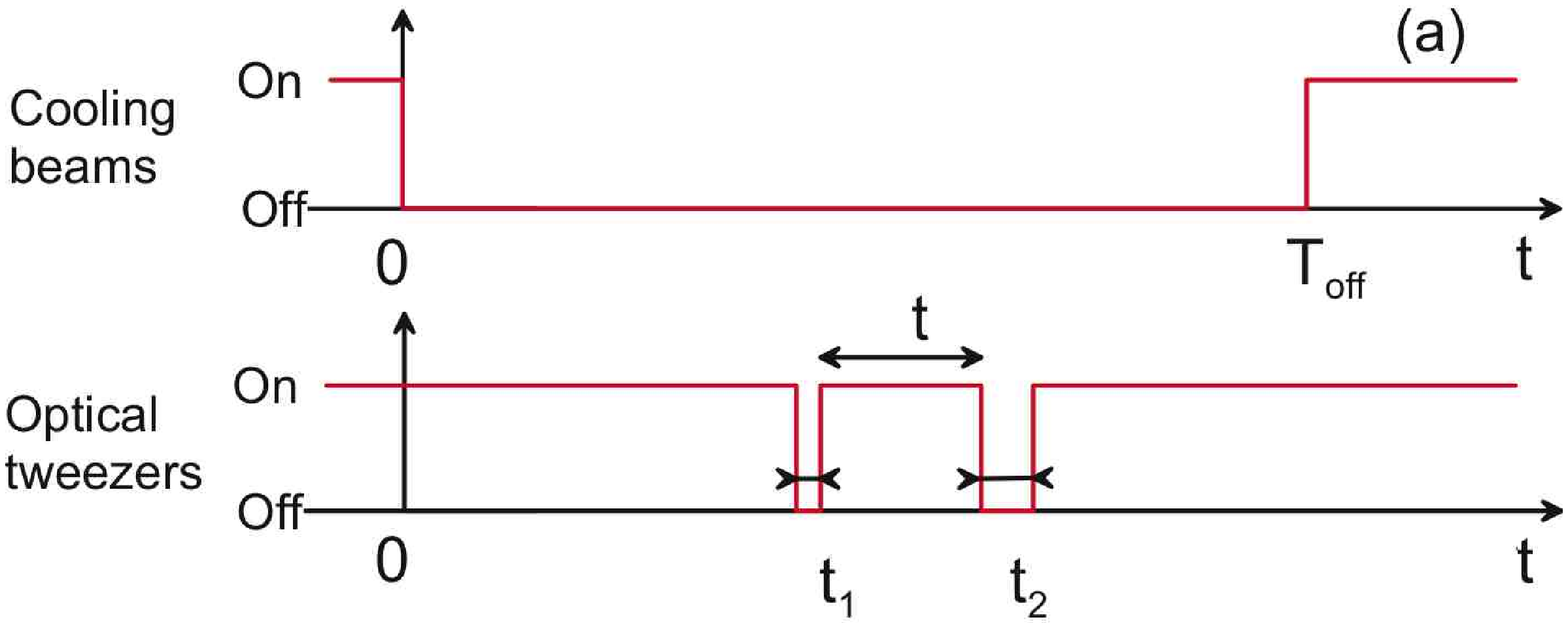}
\includegraphics[width=8cm]{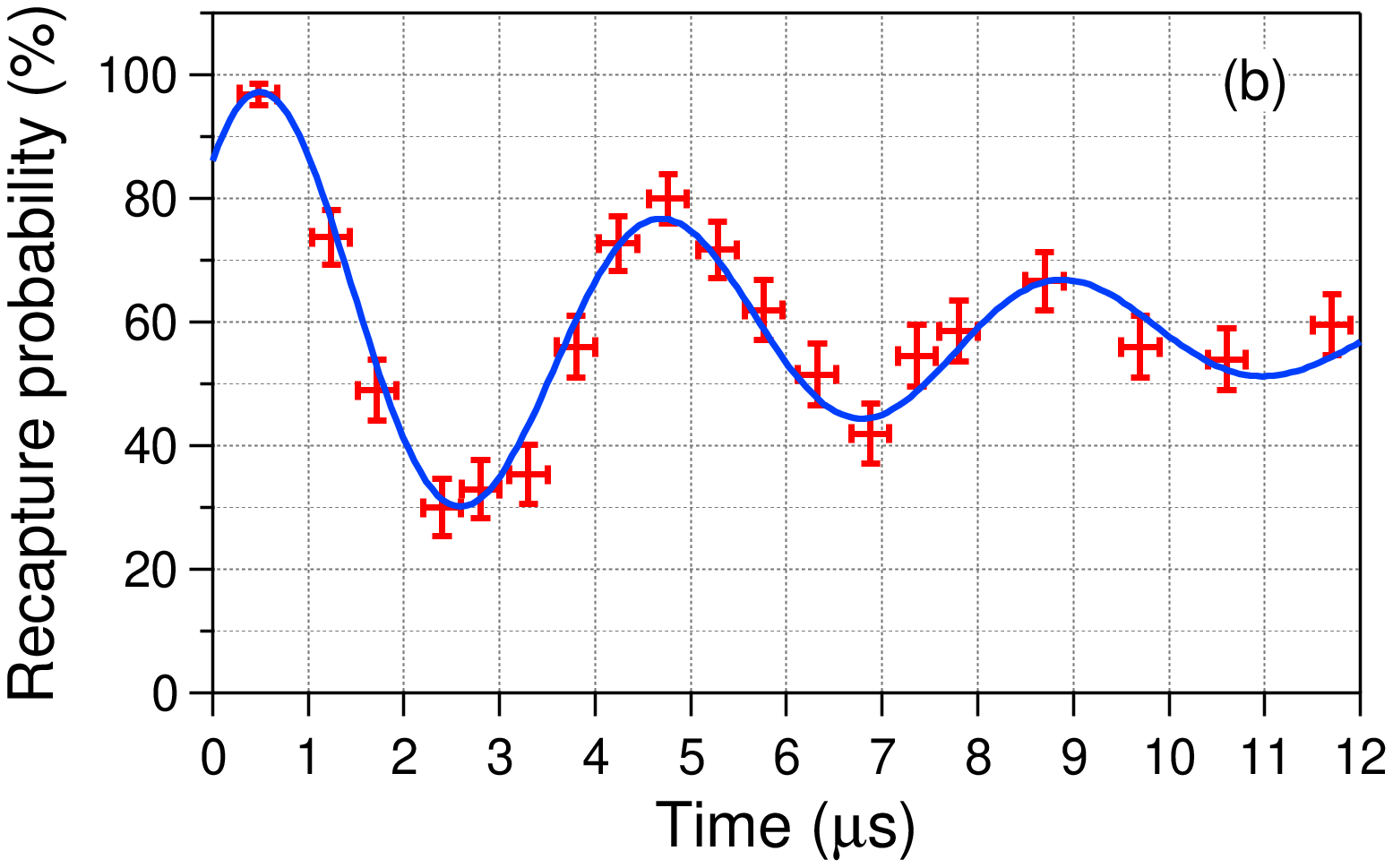}
\caption{(Color Online) Oscillations of a single atom in the dipole trap. (a) shows the time sequence used for this measurement. Dipole trap is
switched off twice, during $\delta t_1\sim1.3~\mu$s and $\delta t_2\sim6.2~\mu$s. In the time interval $\Delta t$ separating these two ``off''
pulses, the dipole trap is switched on again. (b) shows the probability to keep the atom after this time sequence. Each point corresponds to 100
successful events (i.e. with one single atom at the beginning of the sequence) for a given time delay $\Delta t$. Error bars are statistical. Solid
line is a damped sine fit to the data, showing that the atom oscillates with frequency $\omega_r/2\pi=119\pm3$~kHz.} \label{oscillation_frequency}
\end{figure}

In order to measure the oscillation frequency $\omega_r$, we followed the procedure described in~\cite{Reymond2003,DarquieThese2005}. Once a single
atom is loaded in the trap, the cooling beams are switched off for a time $T_{\rm off}=50$~ms. Meanwhile, the dipole trap beam is switched off and on
twice, as shown in the time sequence of Fig.~\ref{oscillation_frequency}(a). The first ``off" pulse (duration $\delta t_1$) increases the initial
amplitude of the oscillation, once the trap is turned back on (see details below). After a variable period of time $\Delta t$, we switch the
potential off again for a fixed duration $\delta t_2$. After this time, we turn the molasses beams on again and determine whether the atom is still
present or not. We then repeat this procedure on about 100 atoms, and measure the probability to keep the atom at the end of this sequence for
various $\Delta t$. We obtain the curve shown in Fig.~\ref{oscillation_frequency}(b).

This curve shows oscillations that we can understand in the following way. Assume the atom is oscillating in the trap. If the atom reaches the bottom
of the trap when the second ``off" pulse occurs, it will most likely leave the trap during the time $\delta t_2$, because its velocity is maximal at
this point. Alternatively, if the atom reaches the apogee of its oscillation when the second ``off'' pulse occurs, it will most likely be recaptured
in the trap after time $\delta t_2$, because its velocity is null at the apogee. Due to the symmetry of the motion, the probability of keeping the
atom oscillates at twice the oscillation frequency when $\Delta t$ is varied. The recapture probability, which we measure when the cooling beams are
switched on again (i.e. at $t=T_{\rm off}$), also depends on duration $\delta t_2$, which is adjusted to optimize the contrast at the beginning of
the oscillation.
\begin{figure}
\includegraphics[width=8cm]{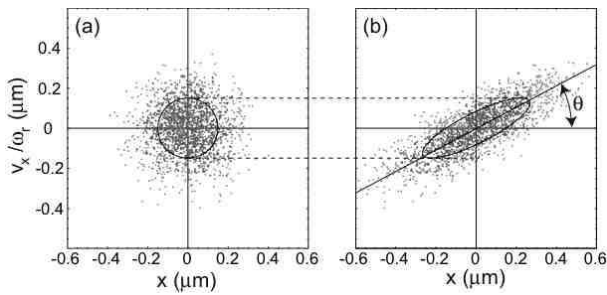}
\caption{Monte Carlo simulation of the evolution of 2000 atoms in the radial plane of the tweezer, using phase space representation. (a) The atoms
are distributed with an initial symmetric gaussian distribution before the first ``off" period of duration $\delta t_{1}$. Circle represents one
standard deviation of the distribution. (b) The distribution evolves towards an ellipse with angle $\theta$ during the free flight of duration
$\delta t_{1}$. As a consequence, the atoms oscillate in phase when they are recaptured in the trap.} \label{rephasing}
\end{figure}

The role of the first ``off'' period is crucial in this measurement. Since each data point shown in Fig.~\ref{oscillation_frequency}(a) is averaged
over 100 single atoms, all these atoms must oscillate in the trap with the same phase if one wants to see any oscillations at all. The first pulse
precisely fulfills this function, as explained in Fig.~\ref{rephasing}. In this figure, we consider the phase space defined by the position $x$ and
the velocity $v_x$ in the radial plane, assuming a harmonic motion with the frequency $\omega_r$. We suppose that the initial distribution of atoms
follows a gaussian distribution, both in position and velocity. The width of this distribution in the phase space $(x, v_x/\omega_r)$ depends on the
mean energy of the atoms. During the free flight of duration $\delta t_{1}$, the initial isotropic distribution evolves towards an elliptical
distribution (Fig.~\ref{rephasing}), because the velocities of the atoms remain constant when the trap is switched off. In phase space, the ellipse
is pulled along the $x$ axis and its long axis makes an angle $\theta$ that decreases with time~: the longer the duration $\delta t_{1}$, the
``flatter'' the ellipse, keeping a constant area in phase space. For an oscillation frequency around 119~kHz and $\delta t_{1}= 1.3~\mu$s, we
calculate that the angle of the ellipse is $\theta\simeq32^{\circ}$ and the ratio between the lengths of the long axis and the small axis is $\sim
2.6$. When the trap is turned back on at the end of the free flight period, the atoms that are still in the trap oscillate with a larger amplitude,
and they are now almost all in phase with each other. We note that for the chosen  $\delta t_{1} $, only  the radial oscillation frequency is
excited, because along the longitudinal  axis  the free expansion remains quite small compared  to the initial size of the atomic ``cloud".

We fit the data shown in figure~\ref{oscillation_frequency}(b) with a damped sine function and measure a radial oscillation frequency
$\omega_r/2\pi=119\pm3$~kHz. The damping of the oscillation is attributed  to the anharmonicity of the trap potential explored by the atom after the
first ``off" period~\cite{Reymond2003,DarquieThese2005}. Since the trapping beam power was measured to be $P_{\rm trap}=5.6\pm0.1$~mW and assuming a
gaussian intensity distribution at the focal point of the trap beam, we infer a beam waist $w_0=1.03\pm0.01~\mu$m. From this result we calculate a
trap depth $U_0=1.5$~mK and a longitudinal oscillation frequency $\omega_z/2\pi=22$~kHz. The waist $w_0$ extracted from the data is in good agreement
with the value of $0.9\pm0.2~\mu$m, obtained by imaging a single atom on a CCD camera, as presented in Sec.~\ref{section:single_atom_detection}.

\section{Conclusion}
In conclusion, we have demonstrated a diffraction limited optical system with a large numerical aperture ($\rm{N.A.}=0.5$) that acts as a sharp
optical tweezer used to trap a single atom. The large collection efficiency of this same lens allows single atom detection with confidence better
than $99\%$ within 10~ms, and resolves atoms at the microscopic level. The resolution is good enough that we can resolve two atoms trapped in two
tweezers separated by less than $2~\mu$m. This system is based on low cost commercial lenses and is relatively tolerant to small misalignments. It
also provides a large field of view ($\pm25~\mu$m) and a large spectral range over which it remains diffraction limited. We believe that this system
is thus a valuable tool for experiments manipulating and observing single particles. It should be useful for applications in quantum computing using
neutral atoms and also for addressing strings of individual ions.

\hskip 1cm

\begin{acknowledgments}
We acknowledge support from the Integrated Project `SCALA' which is part of the European IST/FET/QIPC Program, from Institut Francilien des Atomes
Froids (IFRAF), and from ARDA/DTO. Y.~R.~P.~Sortais and H.~Marion were supported by IFRAF and CNRS fellowships, and A.M.~Lance also by IFRAF.
\end{acknowledgments}


\begin{thebibliography}{32}
\bibitem{numerical_aperture} The numerical aperture is $\textrm{N.A.}=n~\textrm{sin}~\alpha$ where $n$ is the refraction index of the medium
in which the lens is working, and $\alpha$ is half of the angular aperture of the lens. Here $n=1$ since our lens operates under vacuum, and
$\alpha=30^\circ$.
\bibitem{achro} This does not mean that the optics is achromatic~: the images at different wavelengths are in different planes outside the vacuum
chamber, but this is not a limitation since they correspond to different light beams.
\bibitem{IOTA_non_endorsment} Mention of a specific company does not constitute an endorsement by Institut d'Optique Graduate School.
\bibitem{footnote_lens_baking}The vacuum chamber was baked up to $130^\circ$ C for two days. The resulting pressure is less than $10^{-8}$~Pa.
\bibitem{MTF} The incoherent optical modulation transfer function is the 2D-Fourier
transform of the energy point spread function~\cite{Born_Wolf}. The latter being normalized to unity flux, the MTF equals 1 for null spatial
frequency.
\end{thebibliography}
\end{document}